# NanoInfoBio: A case-study in interdisciplinary research[1]

Naomi Jacobs and Martyn Amos

Manchester Metropolitan University

Corresponding author: m.amos@mmu.ac.uk

## 1. Introduction

The complex challenges facing 21st century society will require solutions that transcend disciplinary boundaries (National Academies, 2004). The convergence of informatics, engineering and biotechnology is widely predicted to lie at the heart of the next technological revolution (Carlson, 2008). Interdisciplinary science and technology has the potential to fundamentally transform healthcare, agriculture, energy, security, environmental science and many other areas of pressing concern (Endy, 2005; May, 2009).

Through reconciliation of knowledge across different disciplines, new and innovative forms of research may be stimulated both across and within disciplines (Frost and Jean, 2003, Wilson, 1998, cited in Rhoten, 2004). By encouraging interaction and exchange, the effective creation of new knowledge occurs via the 'cross-fertilisation of fields' (Crane, 1972 cited in Sanz-Menendez et al., 2001). 'Breakthrough' research is more likely to result from combining specialists and ideas from different areas (Carayol and Nguyen Thi, 2004). There may also be financial incentives for undertaking joint research, as novel funding pathways are becoming increasingly available to support such projects. (Cech and Rubin, 2004; Rhoten, 2004; Tadmor and Tidor, 2005).

It has long been recognised that traditional disciplinary boundaries can be limiting, and that these lines of demarcation often create artificial barriers that restrict the type of questions that can be asked (Frost and Jean, 2003). Staying within established boundaries fosters the development of unique worldviews, perceptions, and ways of framing knowledge (Kincheloe, 2001), which are all valuable and necessary. However, communication with those outside the

---

group may become more difficult as a result. Individual researchers may seek to undertake interdisciplinary research (IDR) in order to overcome such limitations (Rhoten, 2004). The complexities of interdisciplinarity, however, are still poorly understood. Whilst IDR appears to have clear benefits, its implementation can offer significant challenges (Kafatos, 1998).

Attention has recently focused on overcoming these challenges, and notable successes include the creation of new interdisciplinary courses, research centres and programmes (Eagan et al., 2002; Aboelela et al., 2007), as well as new policies and funding structures at institutional, national and international levels. In what follows we use an established categorisation of impediments to examine different types of barrier and then illustrate how they have been addressed in the context of an existing research project.

## 2. Case study: the NanoInfoBio project

The NanoInfoBio (NIB) project[2] at Manchester Metropolitan University was funded by the Engineering and Physical Sciences Research Council (UK) for a period of 27 months, under their Bridging the Gaps programme. This programme aims to foster interdisciplinary research within UK higher education institutions by means of innovative and flexible research support. The NIB project brought together computer scientists, biologists, engineers, chemists, mathematicians and health scientists to work on problems as diverse as the fungal deterioration of film stock, wound repair using nanoparticles, and visual tracking of muscle contraction.

The specific objectives of NIB were as follows:

- Encourage serendipity: encourage "happy accidents".
- "Grow our own" researchers: create a sustainable research environment by developing students as researchers.
- Minimise barriers: remove impediments to effective inter-disciplinary work.

---

[2] See the project website at http://www.nanoinfobio.org for further details.

These have been achieved using a variety of methodologies, including seed-corn funding, support for interdisciplinary activities, and a number of initiatives designed to encourage a cultural shift towards more collaborative working across the Faculty of Science and Engineering. We now focus on how the third objective (Minimise barriers) was achieved.

Siedlok and Hibbert (2009) list a variety of factors that can contribute to the failure of IDR, grouped into four categories:

(1) Disciplinary (e.g. cultural barriers)
(2) Personal (e.g. lack of experience, time constraints)
(3) Institutional (e.g. funding schemes, career constraints, authorship/patenting issues)
(4) Procedural (e.g. lack of access to evaluation tools).

NIB took a strategic approach to overcoming barriers to IDR. We now examine each of the categories listed above, propose ways in which they might be overcome, and then describe the implementation of these methods in the context of the NIB project.

## 2. Disciplinary barriers

3.1 The nature of disciplinary barriers

Boundaries can lead to the emergence of particular styles of thinking and approaches to research within a discipline. It may be argued that there are very good reasons why disciplines are the preferred/traditional method for delineating academic research. Boundaries set by disciplines define the parameters and scope of new information to be considered, whereas an "open-ended" framework could potentially overwhelm researchers (Bruce et al., 2004).

Researchers are often unwilling to move outside a personal perspective of their own discipline, a view which may have developed over the course of many years (Gooch, 2005). Participation in IDR may also be resisted due to a perceived tendency of individuals to discriminate against people from outside their self-defined category (the 'in-group') (Fay et al., 2006).

Even if an initial reluctance to move outside disciplinary barriers is overcome, there remain issues with combining the study methods of different disciplines. A lack of shared mental models, common language and assumptions may prove problematic, particularly when participants in a collaborative team have a particularly strong affiliation to their own groups. One of the reasons disciplines assist with concentrated study is that they create a shared framework of thought, through which all members of the discipline may share a continued cohesive frame of reference (Lattuca, 2002). However, if this framework is removed, or if two or more are merged (as is often the case in IDR), the lack of clear structures and rules for conducting research may prove a barrier to effective research (Bruce et al., 2004). For IDR to succeed, dialogue and common ground must be established and maintained between those who have historically sought to distance themselves from disciplines beyond their own.

Frost and Jean (2003) similarly note that disciplines (and institutions) each have their own patterns of attitudes, meanings, symbols and behaviours, and that the thoughts and behaviours of discipline members are influenced by the "knowledge traditions" in which they reside. These include categories of thoughts, common vocabularies and codes of conduct.

> Building the wheel is difficult enough when one person builds the wheel; now try to have three to five people working on the wheel with different tools and different ideas about what kind of bike it will go on.

Here, Morse et al. (2007, p.9) quote a participant in their study using an analogy to describe the challenges they encountered.

There exists a significant body of literature devoted to discussing communication as a barrier to IDR (e.g., Wear, 1999). Disciplines create their own particular vocabularies in order to define and describe terms. These lexicons may not be transferable to other disciplines, and can cause comprehension issues even if the topic under discussion is simple and unambiguous (Jeffrey 2003, Massey et al., 2006). Researchers undertaking IDR have reported issues of this nature, where the same word can have very different meanings in the 'languages' of different disciplines (Bruce et al., 2004). For example, Tadmor and Tidor (2005) describe how the concept of a 'model' differs greatly between biology and engineering, and that this difference must be addressed before effective collaborative work can be carried out. Pickett et al. (1999) note that issues of communication can arise even when terms are understood, due to differences in context, and because of assumptions. Therefore a 'common meaning' is just as important as a common language.

3.2 Overcoming disciplinary barriers

Fay et al. (2006) suggest ways in which projects can both avoid the previously-discussed discipline-based misinterpretations and facilitate the development of shared mental models and common ground. They emphasise the importance of building a cohesive project 'group' via methods such as frequent non-project-related interactions, and the creation of high level goals which are shared and supported by the entire project team. Communication issues arising from disciplinary differences must be addressed by implementing both formal and informal communication strategies (Morse et al. 2007). Others also highlight the importance of space for 'social time', and that strategies for overcoming disciplinary differences require shared space:

> [There is little focus on] the creation of social spaces such as occasions, events, networks, hierarchies, roles and routines that provide opportunities for people to transform disciplinary boundaries, in addition to the creation of common physical spaces (e.g., office location and layout, physical resources, shared seminar rooms, foyers) to foster interdisciplinary objectives. Scott and Hofmeyer (2007, p. 492).

A study by Lee et al. (2010) establishes a correlation between physical co-location and impact of research (at least, in the biomedical sciences), further supporting the argument in favour of providing physical space for collaboration. Bruce et al. (2004) note that in order to overcome issues which may arise from the removal of disciplinary boundaries, defined project boundaries must still be set in order to provide structure and focus for researchers. However, these should be fluid and permeable enough to allow for re-engagement with discipline-based expertise that might initially be outside the project scope, should it be required later.

3.3 Implementation of methods for overcoming disciplinary barriers

The main methods identified in the previous section for overcoming disciplinary barriers were (1) shared, common language; (2) creation of shared goals; (3) physical (perhaps temporary) co-location.

Within NIB we implemented the following strategies for addressing these:

(1) All requests for funding were assessed by an inter-disciplinary team, and applicants were required to phrase their proposals in language aimed at the educated non-specialist. This prevented the over-use of jargon, and encouraged participants to think about fundamental assumptions they held about their work.
(2) Small-scale seed funding was offered for well-defined projects, with short-term objectives and end dates, and the requirement that grants be held by an interdisciplinary team.
(3) Informal meetings were arranged to allow for nonspecific discussion, and to allow bonds to form between individuals and groups who may not otherwise interact. These were often held off-campus, in order to prevent distractions caused by the everyday working environment.

## 4. Personal barriers

4.1 The nature of personal barriers

In van Rijnsoever and Hessels (2010), the authors investigate which personal characteristics are most closely-associated with successful interdisciplinary work. Individuals possessing certain personality traits appear to be more suited than others to interdisciplinary work, one example cited being a "concern with applications". Bruce et al., (2004) list some qualities which are thought to be related to success as a manager or co-ordinator of interdisciplinary research, including willingness to accept alternative methodologies, the ability to learn rapidly, good leadership skills, an interest in "real-world" problems, and a clear vision of the project and what it is trying to achieve. They report a commonly-held view that such personality factors are at least as important as a participant's discipline or specialisation.

There may also be key skills for IDR that are different to those required for discipline-focused studies. Palmer (1999) discusses different strategies of information gathering and knowledge acquisition, and asserts that a broader, more expansive approach to reviewing available literature ('information probing') is more applicable to interdisciplinary research. The aggregation of an appropriate set of project-relevant skills requires a certain level of diversity across participants, as well as complementarities of skills, and a common core of understanding regarding the central problem the project addresses. While these factors are important to all collaborative projects, they may be particularly elusive in those where the participants originate in different disciplines.

In an analysis of the factors affecting collaborative working, Amabilie et al. (2001) list three categories of characteristics which impact on the success of collaborative projects: (1) collaboration skills; (2) project-relevant skills; (3) attitudes and motivation. The authors argue that the most important positive factor in the last category is trust, which is characterised by both an absence of hidden agendas and the existence of mutual respect in the collaborative group. Rowe (2003) identifies investigator-specific factors, which appear in the most part to fall under "attitudes". The factors listed include passion for the work, mutual respect between scientists in the team, complementary skills and knowledge, and the ability to develop a common language. From a negative perspective, personal disputes may arise over matters such as authorship, patenting and data ownership, which can obstruct effective collaborative research (Naiman, 1999, Gooch 2005).

4.2 Overcoming personal barriers

Jeffrey (2003) observes that the skills required for successful IDR are different from those necessary for individual research, and effort must be made to acquire the appropriate skills (such as the ability to integrate different perspectives and communicate effectively with researchers from other disciplines). As discussed above, there exist well-defined personal behaviours and attitudes which can affect the success of IDR (Aboelela et al., 2007, Bruce et al. 2004). The ability of collaborators to meet face-to-face is highlighted by several authors as being a key principle of successful collaborative research (e.g. Rowe, 2003; Maton et al., 2006), and it appears likely that personal interaction is an effective way to overcome many of the barriers discussed here.

It may be necessary for institutions or departments to provide support to help overcome personal barriers, especially to researchers at an early career stage. Additional support for this argument comes from the findings of Bruce et al. (2004), who observe that the degree of interdisciplinarity within a project group appears to increase over time, and increases in line with the learning experience of those involved. This implies that IDR-relevant skills can be acquired, perhaps by continued exposure to alternative disciplinary cultures and attitudes.

Putting into place graduate and postgraduate programmes that provide training in and exposure to IDR will foster these skills, and provide a strong basis from which to develop future interdisciplinary infrastructure (Tadmor and Tidor, 2005). Focussing on nurturing appropriate skills in early career researchers can be crucial in developing long-term IDR expertise.

4.3 Implementation of methods for overcoming personal barriers

The main methods identified in the previous section for overcoming personal barriers were (1) face-to-face contact, (2) support for early career researchers.

Within NIB we implemented the following strategies for addressing these:

(1) Face-to-face interaction was encouraged through informal meetings and sandpit events.
(2) Early career researchers were particularly encouraged to apply for funding, with career stage being a factor in funding decisions for several calls (that is, junior researchers were prioritised).
(3) The involvement of undergraduates and Master's students in funded research projects was financially supported (through summer project bursaries) as part of the "grow our own" ethos of the project.
(4) Special events for postgraduate students were organised in order to introduce them to the benefits and opportunities inherent to interdisciplinary research.

# 5. Institutional barriers

5.1 The nature of institutional barriers

Within academia, the prevailing disciplinary-focused structures and general academic culture can often discourage interdisciplinary work, characterising it as "second-class research" (Siedlok and Hibbert, 2009) or as a "distraction" (Shinn, 2006). If reward structures and funding both within and outside the university are based on discipline-based divisions, they may actively discourage those wishing to engage in cross-disciplinary work. Nobel laureate Russell Hulse notes:

> Just setting up interdisciplinary centers at universities doesn't get you where you want to go if you haven't changed the reward system. Goodman et al. (2006, p.1235).

Leshner (2004) notes that, since most universities are organised into discrete departments in order to promote scholarship within their particular 'disciplinary silos', they are not well-positioned to facilitate IDR, and may effectively penalise such work. This is particularly true if the individual disciplines do not regard the interdisciplinary areas of research as appropriate for engagement. For example, Cech and Rubin (2004) cite delays in the development of the now booming area of bioinformatics, mainly due to the fact that it was initially embraced by neither biology nor computer science departments.  Postgraduate students wishing to engage in IDR may encounter problems such as finding a sympathetic supervisor, and having to spend additional time gaining mastery of potentially conflicting disciplines (Golde and Gallagher, 1999). At a more senior level, there exists a lack of understanding that impacts on tenure decisions, based on (for example) the fact that an academic's position in the author list of interdisciplinary publications may not accurately reflect their level of contribution (Cech and Rubin, 2004).

The US National Academies (National Academies, 2004) outline ways to stimulate and encourage IDR, focusing mainly on improvements to institutional structures, and indicating that current systems are unsatisfactory. It reiterates the fact that many institutions claim to support IDR and see its value, but expect staff to take on IDR-related responsibilities as additional duties over and above their usual obligations. Time constraints are often cited by researchers as a significant barrier to successful IDR (see Morse et al., 2007).

5.2 Overcoming institutional barriers

Careful management and planning is crucial to the success of IDR projects. Carayol and Nguyen Thi (2004) find that, while recurrent public funding has no discernible effect on interdisciplinarity, contractual funding from private and public sources has significant positive effects. Organisational arrangements can also either support or obstruct IDR; therefore changes may need to be made at the institutional level in order facilitate IDR.

Detailed suggestions for the forms these changes might take are outlined in a report by the European Union Research and Advisory Board (EURAB, 2004). The main recommendations of the report are divided into five categories, the first of which focuses on ways to avoid unnecessary administrative barriers. The report suggests that a balance is needed between highly-specific funding mechanisms, and the ability to fund broad IDR. It also suggests that departmental and faculty divisions, and the associated employment procedures, should be examined to ensure that they do not create barriers to IDR. The development of shared research facilities is also discussed, and the provision of intra- and inter-institutional access to any newly funded major research infrastructure and facilities is suggested. The funding and management of IDR is also examined, and the authors make specific suggestions for changes to funding procedures and resource allocation methods to encourage interdisciplinary work, as well as proposing methods for both the dissemination of good practice, and the identification of new research fields.

Suggested changes to institutional and funding structures are described in further detail in a report by the US National Academy of Sciences (National Academies, 2004). This includes recommendations to funding organisations, suggesting that they should consider in their programmes and processes the unique challenges of IDR, with respect to risk, organisational mode and time. Recommendations are also offered to academic institutions, highlighting a need to remove barriers to IDR, with several illustrative examples from institutions that have successfully enabled necessary changes. The specific recommendations are detailed and extensive, but include the following:

(1) Streamline fair and equitable budgeting procedures across department or school lines, to allocate resources to interdisciplinary units outside departments or schools.
(2) Allocate research space to projects, as well as to departments.
(3) Deploy a substantial fraction of flexible resources - such as seed money, support staff, and space – to support IDR. (National Academies, 2004)

Because some of the issues described above result from a lack of understanding at the senior management levels of institutions (for example, lack of additional time provisions for IDR), the key suggestion is that organisational learning and change is required to successfully foster IDR, as well as individuals developing new skills and approaches (Lattuca 2002).

5.3 Implementation of methods for overcoming institutional barriers

The main methods identified in the previous section for overcoming institutional barriers were (1) support cross-disciplinary funding, focussing on specific projects, (2) provide appropriate infrastructure support, (3) encourage high-level institutional awareness of IDR and its benefits.

Within NIB we implemented the following strategies for addressing these:

(1) Provision of small-scale, seed-corn funding to support start-up projects, as well as larger amounts for further development of successful ideas. This was particularly effective when used to "buy" short periods of time for postgraduate students and/or technical staff, consumables and small pieces of equipment. We initially proposed also using these funds to support teaching buy-out, whereby academic staff could have portions of their teaching covered by adjuncts / associate lecturers. However, this was less successful, as the administrative overhead involved in arranging buy-out often made it impractical.
(2) A full-time administrator was appointed to run the programme on a day-to-day basis and deal with purchasing, travel, meetings, etc. Given the right appointment, the administrator can also have significant input into the strategic development of IDR within an institution.
(3) The project was actively promoted, both within and outside the institution, via public lectures, a dedicated website, blogs, collaborative events and newspaper / magazine articles. This was vital in terms of gaining senior support for the project (at the level of Deans and above), as well as providing positive publicity for the institution. The project brand became increasingly important, and a "corporate identity" (logo, colour scheme, NanoInfoBio phrase) was used throughout in order to cement awareness.

## 6. Procedural barriers

6.1 The nature of procedural barriers

Siedlok and Hibbert (2009) suggest that IDR can often develop in an unsystematic manner. Although such "emergent" research can be very productive, it may also result in a lack of structured processes for its management and practice, and ultimately lead to conflict.

Amabile et al. (2001) provide several examples of process-based issues in collaborative projects. These include frustration with initial project meetings that lack active discussion and decision making. The effective use of meeting facilitation skills at subsequent gatherings appears to address this issue. It may also be the case that different administrative processes are used in different departments or disciplines:

> Determining the necessary criteria for accessing…resources was often a confusing and complex endeavour because our departments and colleges all had different policies and procedures regarding resource acquisition and allocation that often conflicted with each other. Koch et al. (2005, p.371).

6.2 Overcoming procedural barriers

Process-based barriers can be overcome by appropriate planning to ensure that potential issues are identified at the start of an interdisciplinary project, and by introducing measures to address possible problems. Morse et al. (2007) suggest that an accountability strategy, setting out interdisciplinary team timelines, requirements and responsibilities, is essential for integrated working.

Rhoten (2004) observes that the most successful IDR projects allow researchers to freely enter and exit short-term collaborations. It is important to avoid fixing long-term collaborations at the start of the project, based on a 'laundry list' of affiliates chosen to fill a specific number of positions rather than for their skills. Researchers report that when they are free to move between collaborations they make more progress with interdisciplinary projects and have greater overall satisfaction in their professional lives. Rhoten also finds that the size of interdisciplinary centres and networks is key to their success, and that small centres (or small bounded networks within large centres, with fewer than 20 affiliates) are found to generate more knowledge creating connections than medium/large centres.

The importance of research group size is also highlighted by Cech and Rubin (2005), who describe two examples of highly successful interdisciplinary research organisations (the Medical Research Council Laboratory of Molecular Biology [MRC LMB], and the former AT&T Bell Laboratories). They emphasise the fact that at both these centres, individual research groups are generally composed of fewer than six individuals, and that this small group size is considered a critical factor in promoting effective collaboration. To undertake larger projects, small groups would themselves collaborate.

Jeffrey (2003) highlights the important role facilitators can play in interdisciplinary projects, and describes ways in which facilitation assists in the process. These include maintaining a focus on collaborative aspects of the project, taking decisions which may be seen as unpopular without damaging the collaboration, and assisting in the development of common vocabularies.

While this suggests that planned learning is necessary to support the development of interdisciplinary teams, work by Lattuca (2002) indicates that informal patterns of learning may also be relevant. Those who participate in serendipitous interdisciplinary collaboration often find that this provides skills useful for further collaborative work, and that learning occurs 'in situ'. Participants in the study by Bruce et al. (2004) regarded the best collaborations as those which built upon existing links and contacts, supplemented by new contacts initiated either by direct contact with one of the members, word-of-mouth suggestions or via other informal contacts.

Successful interdisciplinary research projects must report methodologies and processes used, so that they can be replicated by subsequent projects. Although the nature of collaboration will differ, by necessity, for each collaborative project, explicitness allows for ease of evaluation and future learning (Robertson et al., 2003).

6.3 Implementation of methods for overcoming procedural barriers

The main methods identified in the previous section for overcoming procedural barriers were (1) ensure consistent accountability, administrative and reporting processes are in place, (2) keep collaborative groups relatively small, (3) make use of facilitators, where appropriate.

Within NIB we implemented the following strategies for addressing these:

(1) By having a central administrator to oversee the project, there existed a natural first point of contact for administrative queries. This, in turn, made it easier to use consistent administrative processes for purchasing, auditing and project reporting. The administrator ensured a consistent "front" for the project by generating a set of forms (e.g. request for travel funds) for use only within the project, which were then mapped onto central university systems. Although this added an extra layer of complexity, it ensured that a complete and accurate audit trail was in place, and applicants had a uniform experience when requesting funding or other assistance.

(2) Collaborative groups were kept small by virtue of the fact that the individual sums of money available via the project were relatively small. Project teams of size greater than three or four individuals were unusual.

(3) The administrator often served as the project facilitator by bringing together people with common interests via meetings and introductions. Each funded project was required to hold kick-off meetings, and, for the larger projects, regular status update meetings. Where possible, these were attended by the administrator, whose presence was intended to ensure that misunderstandings due to differing uses of specialist terminology were avoided, since it was necessary to describe achievements and objectives in non-specialist terms.

# 8. Conclusions

In this paper we have emphasised the future importance of interdisciplinary research, and outlined some of the challenges to its effective implementation within universities. By describing ways in which a funded project has addressed these barriers, we hope to offer to the community tangible and useful examples of good practice, and contribute to a wider debate on the implementation of cross-disciplinary science and engineering.

# Acknowledgments

The authors gratefully acknowledge the support of the Engineering and Physical Sciences Research Council Bridging the Gaps programme (Grant reference EP/H000291/1).